\documentclass[floatfix,
 reprint,
 superscriptaddress,
 amsmath,amssymb,
 aip,
 jap,
]{revtex4-1}

\usepackage{textcomp}
\usepackage[utf8]{inputenc}
\usepackage{enumitem}

\usepackage{graphicx}

\begin{document}

\title{Patterning perovskite colour converters for AR/VR microdisplays}

\author{Ruairi Baker}
\author{Maria Pervez}
\author{Angus Hawkey}
\author{Nobuya Sakai}
\author{Valérie Berryman-Bousquet}
\author{Bernard Wenger}
\email{bernard.wenger@heliodm.com}
\affiliation{Helio Display Materials, WCFI, Quarry Road, Oxford OX3 8SB, UK}

\date{\today}

\begin{abstract}
Colour conversion offers the clearest path to achieve RGB colours in high resolution microdisplays for AR/VR. With resolutions beyond 5000 ppi (i.e. RGB pitch of ~5 µm), the thickness of the conversion layers is critical for efficiency and manufacturing. Perovskites outperform other conversion materials (quantum dots or phosphors) with their high absorption coefficients for blue light. In this contribution, we show how perovskite materials, engineered for high optical density and colour purity, can be patterned to produce colour converting pixels. We demonstrate patterning using three approaches (lift-off, negative photoresist and dry etch), and discuss their advantages and disadvantages. The results consolidate the choice of perovskites for AR/VR applications by demonstrating their robustness and compatibility with multiple patterning strategies suitable for high resolution microdisplays.
\end{abstract}

\maketitle

\section{Introduction}
MicroLEDs hold the key to the brightness AR microdisplays demand, but achieving vibrant RGB colours remains a significant technical challenge. Colour conversion (CC) is so far the most realistic approach to achieve this target. It combines brightness, efficiency and colour quality without compromising on the performance of the blue microLEDs emitters. Requirements for colour conversion materials in microdisplays are demanding, but perfectly suited to perovskite nanocrystals. Beyond their high conversion efficiency and superior blue light absorption, perovskites are chemically defined, avoiding the inconsistencies tied to nanocrystal size variations in QDs, and are more easily manufactured at convenient temperatures. \\
A key promise of MicroLED displays is ultra-high resolution (5000 ppi and beyond), enabling high image quality for AR/VR headsets. At 5000 ppi (RGB pitch of 5.1 µm) sub-pixels on the order of 1-2 µm are required. While state-of-the-art GaN based blue µLEDs can be effectively manufactured at this scale, such pixel size sets a strong constraint for the absorption power of a colour conversion material deposited on top. If too thick, CC pixels are prone to mechanical instability and substantial losses (and/or cross-talk) due to light mostly escaping from the sides of the pixels. Instability can be mitigated by containment within a bank structure, but it has been shown that absorptive banks were the major loss mechanism when the aspect ratio increases [1]. Therefore, CC materials with highest optical density in the blue are required. Due to their intrinsic absorption properties, Perovskites nanocrystals excel at this task, and, using proprietary synthesis methods and tailoring of nanocrystals surfaces, we achieve OD/µm $>$ 1.0 for both red and green pixels. To our knowledge, this is about 2x higher than best reports for Cd-based quantum dots [2], and more than 5x larger than for InP QDs [3, 4] . This translates into high colour quality (i.e. minimal blue leakage) achieved in 2-3 µm thick pixels without the need of additional colour filters.\\
In this contribution, we discuss different methods to pattern CC pixels based on perovskite nanocrystals and evaluate their potential for application in microdisplays for AR/VR. After briefly discussing the properties of the synthesized perovskites, we introduce a lift-off process suitable for thin pixels, then we present a conventional photolithography method based on a negative photoresist loaded with perovskite nanocrystals, and finally a dry etching approach

\section{Results}
\subsection{Perovskite colour conversion materials}
Green and red perovskite nanocrystals were synthesised using proprietary synthetic routes and surface engineering, and dispersed in toluene at high concentration (400-500 mg/mL). Typical absorption and emission spectra for CsPbX\textsubscript{3} (X = Br and/or I) films are shown in Figure 1. Unlike quantum dots, the peak wavelength of perovskites is mostly defined by their chemical composition and not their size (weak quantum confinement regime). For red perovskites, the ratio of iodide and bromide is finely tuned to adjust the emission peak at the desired wavelength (here 623 nm). Narrow emission peaks are achieved for both colours (FWHM: green 20 nm, red 32 nm), allowing for an ultrawide colour gamut.

\begin{figure}
\includegraphics[width=\columnwidth]{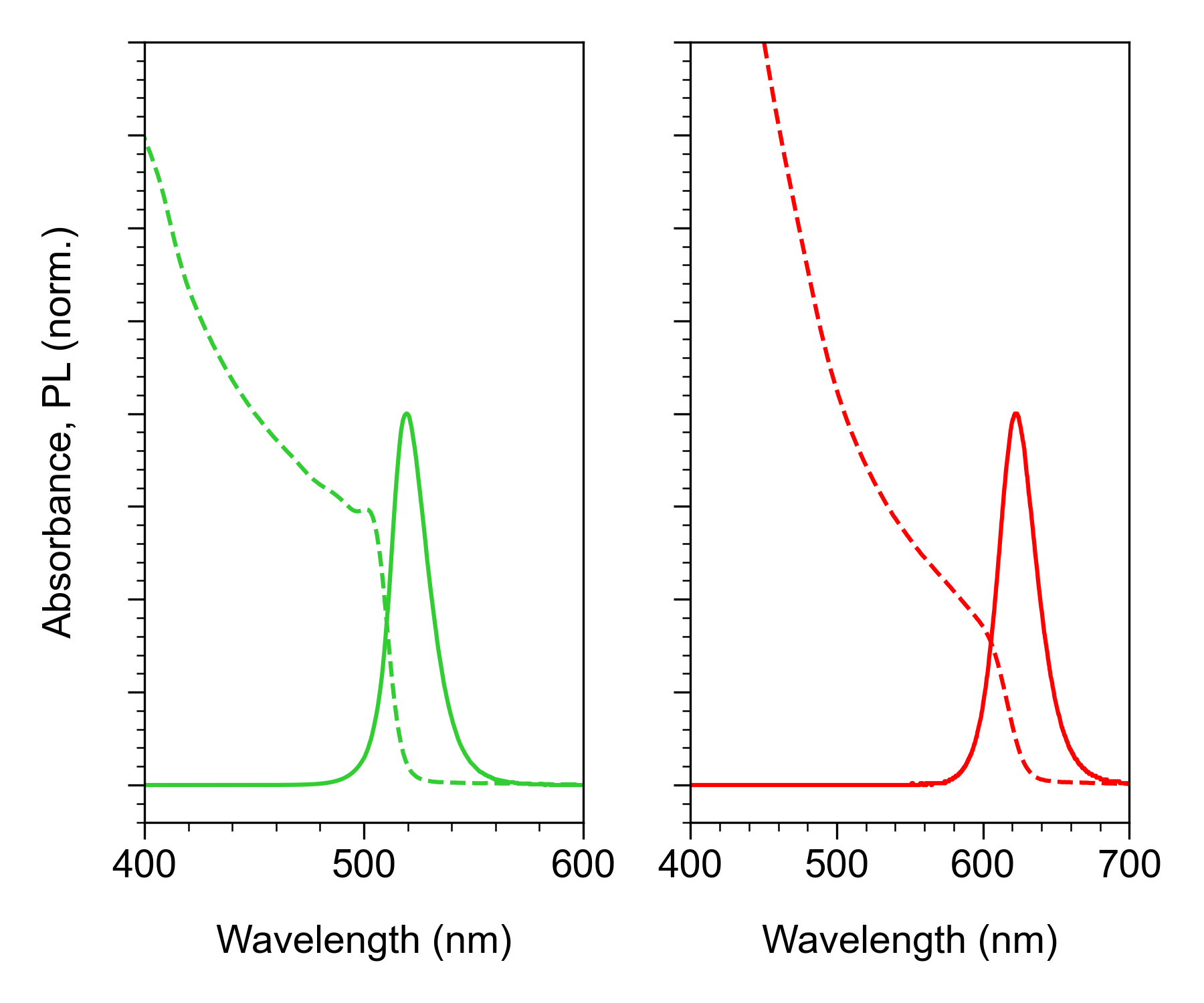}
\caption{\label{fig1}Absorption and photoluminescence spectra of typical green (left) and red (right) perovskite thin films.}
\end{figure}

\subsection{Lift-off process}
For AR/VR applications, where the required resolution is below 10 µm, printing methods are generally not suitable. Advanced methods, such as electrohydrodynamic printing can achieve the desired pixel size but suffer from strong limitations in processing time due to the sequential printing process. Therefore, we focused essentially on microfabrication methods suitable for industrialisation and already available in standard semiconductor fabs. \\
The first method we demonstrate is based on a lift-off approach, where perovskites are uniformly coated  on top of a patterned photoresist that is subsequently stripped away (See Figure 2a). While this method can effectively provide high resolution, it may be challenging to find a photoresist fully compatible with colour conversion materials. For example photoresists are usually removed with organic solvents similar to those used to disperse perovskite nanocrystals and thus will be damaged during step 2a, or the perovskite will be removed with the resist at step 3a (Figure 2a). We found that lift-off was possible using photoresists based on fluorosolvents and fluoropolymers (e.g. Oscor 4020 and developer 103, from Orthogonal inc.). In the absence of cross-linking in the colour conversion material, the lift-off approach is limited to relatively thin pixels ($\sim$0.5-1.0 µm) due to the weak mechanical resistance of small pixels leading to breaking during the lift-off step.\\
\begin{figure}
\includegraphics[width=\columnwidth]{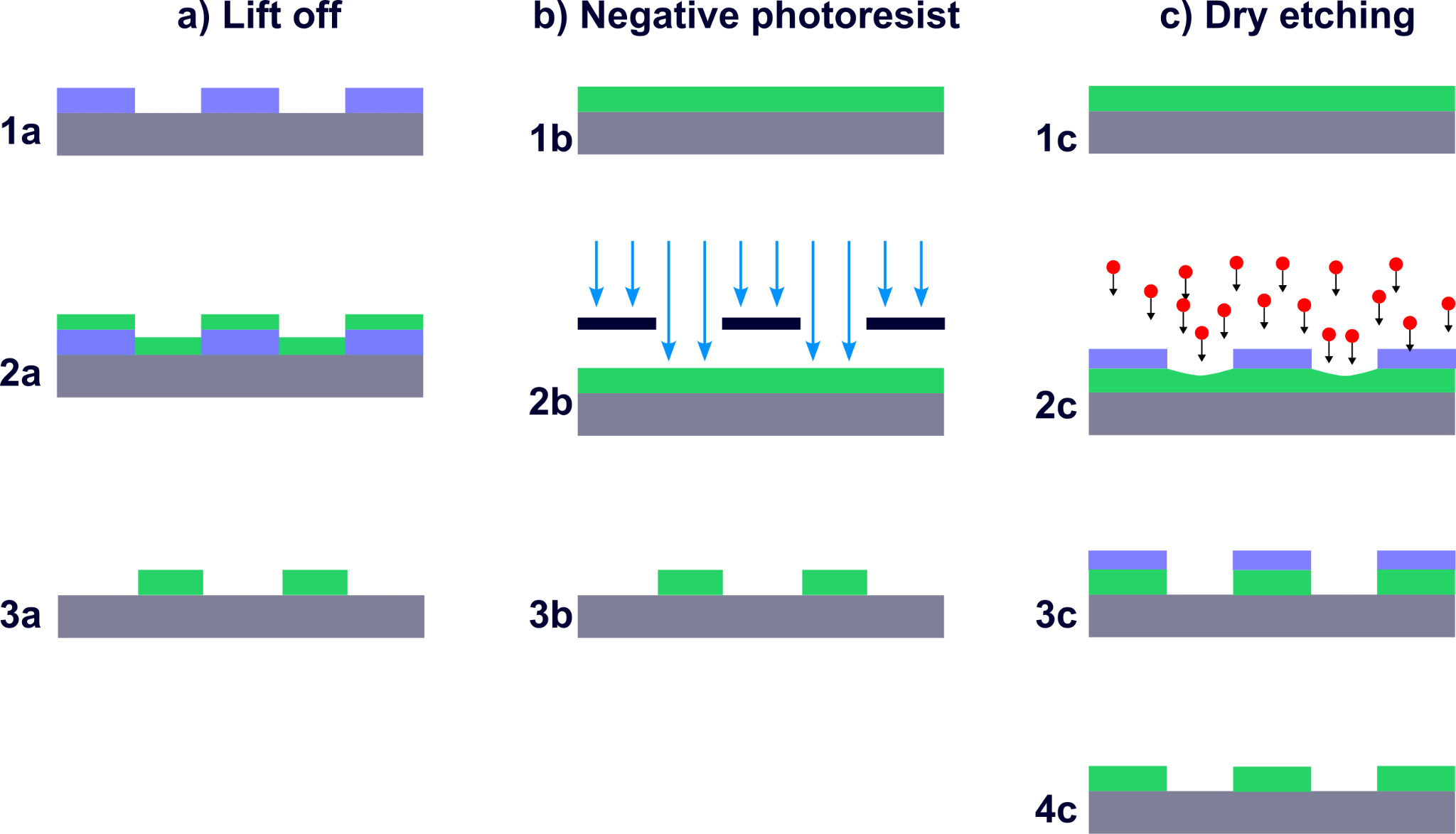}
\caption{\label{fig2}Patterning methods for perovskites colour converters (grey: substrate, green: perovskite material, blue: photoresist). See text for details.}
\end{figure}

\begin{figure}
\includegraphics[width=\columnwidth]{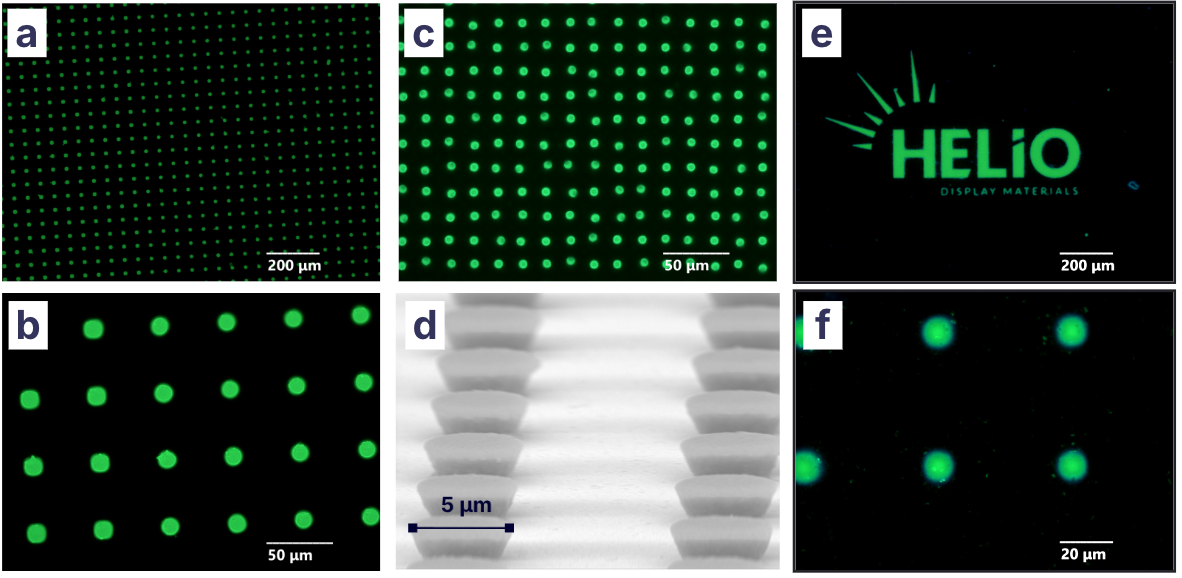}
\caption{\label{fig3}Photoluminescence (a-b, e-f) and SEM images (d) of green perovskites CC pixels patterned via (a, b) lift-off process (pixel diameter ~10 µm), (c, d) negative photoresist (pixel diameter ~5 µm) and (e, f) dry etching (pixel diameter ~10 µm).}
\end{figure}

\subsection{Negative photoresist}
Solutions to address the mechanical stability of the colour conversion pixels include direct UV cross-linking between nanocrystals, or embedding nanocrystals into a UV-curable resin. In other words, formulating a negative photoresist. This approach can be used to increase the thickness of small pixels in the lift-off process (Figure 2a). However, if cross-linking is strong enough the sacrificial photoresist becomes obsolete and a simpler photolithography process can be used (see process in Figure 2b). For AR/VR applications, where the colour conversion pixels should be as thin as possible while minimising blue leakage, key challenges for a negative photoresist are i) maximise loading of perovskite nanocrystals in the cross-linkable resin and ii) ensure cross-linking throughout a highly UV absorbing material. If that latter is not achieved, pixels will not adhere to the surface and will be washed away during development. \\

In this work, we used a combination of measures to achieve these objectives. Firstly, we engineered perovskite nanocrystals to feature cross-linkable functional groups on the surface. Then we developed a highly sensitive resin formulation based on thiol-ene chemistry to achieve good cross-linking at high perovskite loading. And finally, we selected appropriate photoinitators and optimised their concentration. For example, titanocene photoinitiators were particularly efficient due to their high absorption coefficient and broad absorption spectra. In Figure 3c and 3d, we show examples of patterns achieved with a perovskite-containing negative photoresist formulation. Unlike the lift-off method we find that this approach is suitable for thicker pixels up allowing for complete absorption of blue light. The trade-off in optical density per pixel thickness (OD/µm) is illustrated in Figure 4, where the green photoresist has about 25\% less optical density compared to the neat material (without resin). We note that due to the high intrinsic absorption of perovskites both materials achieve OD/µm considerably superior to quantum dot photoresists (QDPR).

\begin{figure}
\includegraphics[width=\columnwidth]{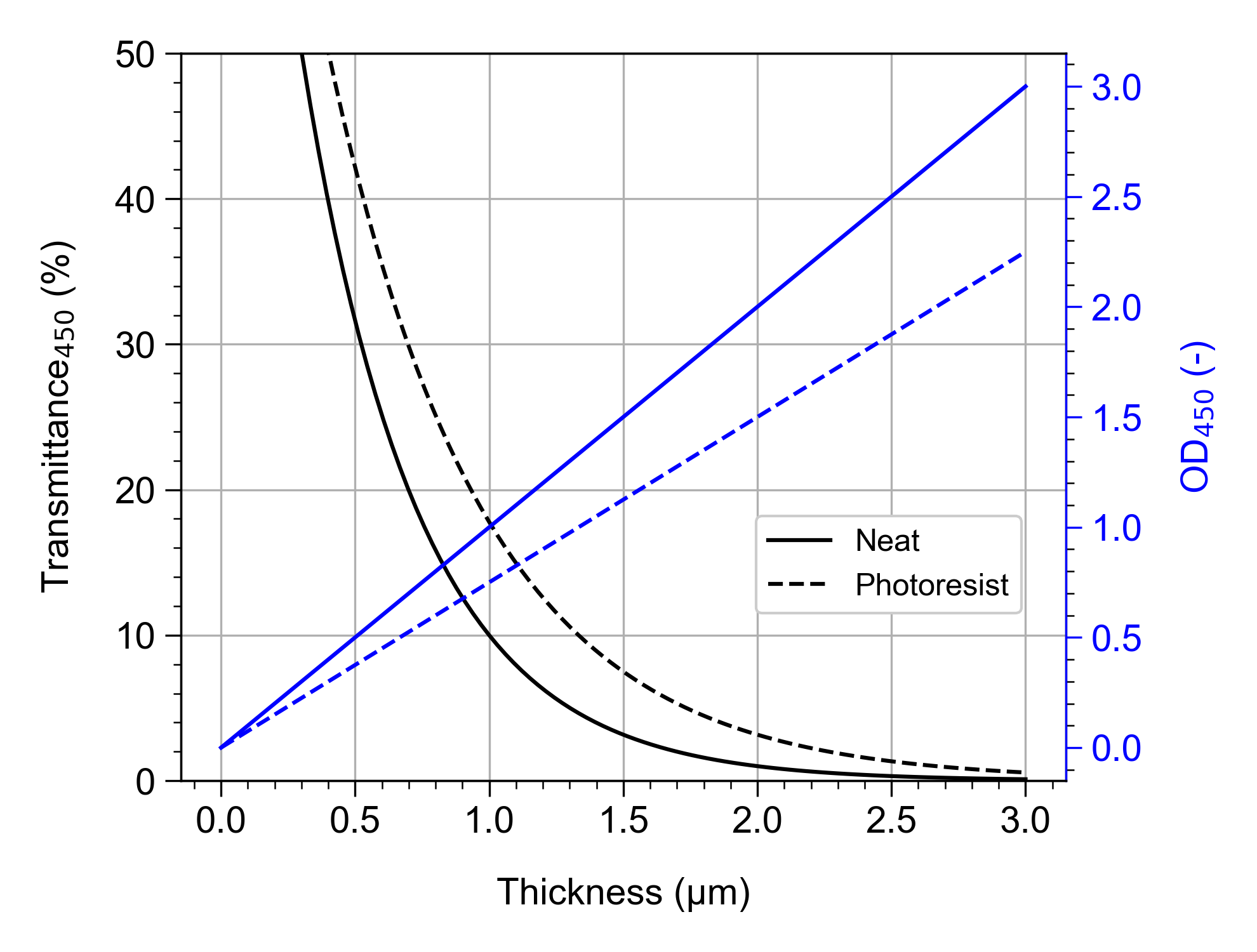}
\caption{\label{fig4}Transmittance and OD at 450 nm for films obtained from green neat and photoresist materials at 450 nm.}
\end{figure}

\subsection{Dry etching}
Dry etching is a common method for patterning hard and soft materials in semiconductor manufacturing, but it has been only occasionally used to pattern colour conversion materials. This may be surprising considering the availability of industrial equipment and its scalability for processes based on wafer-sized substrates such as microdisplays frontplane. A notable exception is a report on the fabrication of neutron detectors [5], using a combination of HBr and Ar plasma to etch CsPbBr$_3$ for neutron detectors. For dry etching processes, the etch mask is a key component that should offer good processing compatibility with the materials to etch and a good etching rate contrast. In their process, Caraveo‐Frescas et al. [5] used a sacrificial chromium hard mask that was consumed during the process. \\
To etch perovskites materials, we found that a Cl\textsubscript{2} plasma could effectively remove µm-thick films. We selected a simplified process using a thick photoresist as etch mask. Similar to the lift-off process we find that we can process a fluorinated photoresist system (Oscor 4020, developer 103) on top of our materials without damage. Since the etch rate contrast is very limited, the thickness ratio between perovskites and photoresist is critical. Using optimised conditions, we achieved homogeneous etching of green perovskite neat materials without residuals (see Figure 3e-f). The thickness of the resulting pixels was limited due to the poor etch rate contrast. We suggest that using a hard mask (such as metals, oxides or nitrides) would significantly enhance the processing window and allow for thicker pixels, without compromising lateral dimensions.

\begin{figure}
\includegraphics[width=\columnwidth]{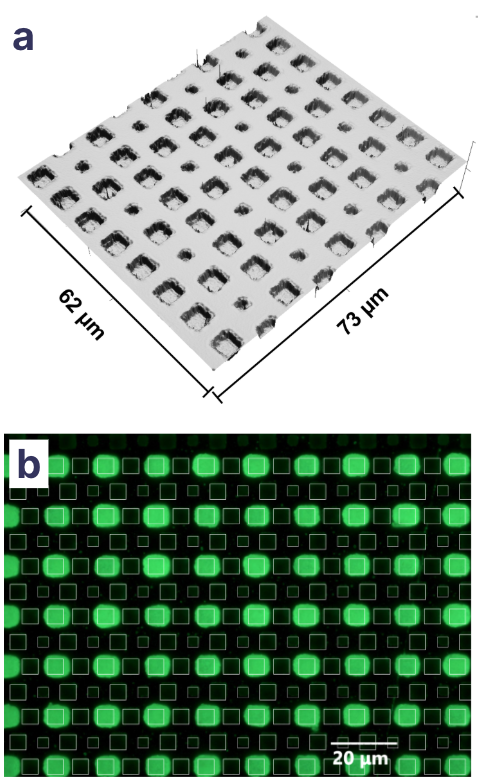}
\caption{\label{fig5}a) Optical profiler image of wells in an aluminium matrix (depth 2.6 µm) and b) photoluminescence micrograph of wells filled with a green perovskite photoresist (position of unfilled wells revealed by white overlay).}
\end{figure}

\subsection{Patterning in matrix}
All three methods described above can provide self-standing pixels with dimensions suitable for high resolution AR/VR microdisplays. However, without pixel isolation structures, the colour quality will be compromised by cross-talk between pixels. Highly absorbing banks (“black matrix”) are commonly used in conventional displays such as LCD or QD-OLED TVs. However, for very small pixels, efficiency losses associated with absorbing side walls are unacceptable. A solution is to use a reflective metallic matrix to contain colour conversion materials. We demonstrate selective filling of an aluminium matrix using the negative photoresist approach. We note however that lift-off and dry etching could also be used. In Figure 5a we show an optical profiler image of a 2.6 µm thick aluminium metal matrix with squared wells. Wells were selectively filled with a green perovskite photoresist (Figure 5b). We observe a slight overfill of the wells due to excess UV exposure.

\section{Discussion and Conclusions}
\begin{table*}[b!]
    \centering
    \begin{tabular}{|p{4cm}|p{4cm}|p{4cm}|}
    \hline
    \textbf{Lift-off} & \textbf{Photoresist} & \textbf{Dry etch} \\
    \hline
    \begin{itemize}
        \item [+] Resolution
        \item [+] No cross-linking
        \item [+] High OD/$\mu$m
    \end{itemize}
    &
    \begin{itemize}
        \item [+] Mechanical stability (cross-linked)
        \item [+] Scalability
    \end{itemize}
    &
    \begin{itemize}
        \item [+] High resolution
        \item [+] No cross-linking
        \item [+] High OD/$\mu$m
        \item [+] Scalability
    \end{itemize}
    \\
    \hline
    \begin{itemize}
        \item [$-$] Solvent compatibility (photoresist vs perovskite)
        \item [$-$] Limited to thin layers.
    \end{itemize}
    &
    \begin{itemize}
        \item [$-$] Lower OD/$\mu$m due to crosslinkers/resin
        \item [$-$] Undercut limiting resolution
        \item [$-$] Damage due to long UV exposure (for thick films with high OD)
    \end{itemize}
    &
    \begin{itemize}
        \item [$-$] Solvent/photoresist compatibility
        \item [$-$] Etch contrast
        \item [$-$] Damage by etching gas/plasma
        \item [$-$] Overetch
    \end{itemize}
    \\
    \hline
    \end{tabular}
    \caption{Pros and cons of patterning method discussed in this paper.}
\end{table*}

In this contribution we demonstrate three different patterning methods suitable for forming small pixels with dimensions suitable for colour conversion in AR/VR microdisplays. These methods have their own advantages and disadvantages that we summarise in Table 1. However, together they show that perovskites materials are versatile and robust materials that can be engineered into formulations suitable for each method. When cross-linking is required, the negative photoresist approach which is analogous to colour filters used in LCDs is preferred. However, with very small pixels the undercut typically associated with negative photoresist may limit adhesion. This is however mitigated when using a host matrix which additionally improves cross-talk and efficiency.

\section{References}
\begin{enumerate}[label={[\arabic*]}]
    \item P. Palomaki, K. Twietmeyer, “26-3: Optical Modeling of Quantum Dot‐OLED (QD‐OLED) Color Conversion”, SID Symp. Dig. Tech. Pap., vol. 53(1), 303-306, 2022. 
    \item C.J. Chena, K.A. Chenb, R.K Chianga. "12‐3: Materialization of Mid‐Resolution Quantum Dot Color Converters on G2. 5 TFT‐LCD Production Line for Micro‐LED Displays." SID Symp. Dig. Tech. Pap., vol. 55(1), pp. 128-131. 2024
    \item I. Jen-La Plante et al., "55‐2: quantum dot color conversion for displays." SID Symp. Dig. Tech. Pap., vol. 54(1), pp. 792-794. 2023.
    \item Y. Kambe et al., "67‐1: High Optical Density, High Efficiency Quantum Dot Photoresist for MicroLED Applications." SID Symp. Dig. Tech. Pap., vol. 56(1), pp. 907-909. 2025.
    \item J.A, Caraveo‐Frescas, M.G. Reyes‐Banda, L. Fernandez‐Izquierdo, M.A Quevedo‐Lopez, “3D microstructured inorganic perovskite materials for thermal neutron detection”, Adv. Mater. Tech., 7(6), p.2100956, 2022
\end{enumerate}

\end{document}